\documentclass[11pt,notoc]{JHEP3}

\usepackage{epsfig}

\usepackage{amsmath}

\def\om{\omega}

\newcommand{\ben}{\begin{equation}}
\newcommand{\een}{\end{equation}}
\newcommand{\bea}{\begin{eqnarray}}
\newcommand{\eea}{\end{eqnarray}}
\newcommand{\bit}{\begin{itemize}}
\newcommand{\eit}{\end{itemize}}

\newcommand{\sn}{\mathrm{sn}}
\newcommand{\cn}{\mathrm{cn}}
\newcommand{\dn}{\mathrm{dn}}

\newcommand{\T}{\ensuremath{{\cal T}}}

\newcommand{\half}{\ensuremath{\frac{1}{2}}}

\newcommand{\ba}{\begin{eqnarray}}
\newcommand{\ea}{\end{eqnarray}}

\title{Perturbations and moduli space dynamics of tachyon kinks}

\author{Mark Hindmarsh and Huiquan Li\\
Department of Physics and Astronomy\\
University of Sussex\\
Brighton BN1 9QH\\
U.K.}

\abstract{The dynamic process of unstable D-branes decaying into
stable ones with one dimension lower can be described by a tachyon
field with a Dirac-Born-Infeld effective action. In this paper we
investigate the fluctuation modes of the tachyon field around a
two-parameter family of static solutions representing an array of
brane-antibrane pairs. Besides a pair of zero modes associated
with the parameters of the solution, and instabilities associated
with annihilation of the brane-antibrane pairs, we find states
corresponding to excitations of the tachyon field around the brane
and in the bulk. In the limit that the brane thickness tends to
zero, the support of the eigenmodes is limited to the brane,
consistent with the idea that propagating tachyon modes drop out
of the spectrum as the tachyon field approaches its ground state.
The zero modes, and other low-lying excited states, show
a 4-fold degeneracy in this limit, which can be identified with 
some of the massless superstring modes in the brane-antibrane system.
Finally, we also discuss the
slow motion of the solution corresponding to the decay process in
the moduli space, finding a trajectory which oscillates
periodically between the unstable D-brane and the brane-antibrane
pairs of one dimension lower.}


\begin{document}

\section{Introduction}
\label{sec:introduction}

In Type II string theories, there are two kinds of branes: stable
or BPS branes, which are supersymmetric and charged, and unstable
or non-BPS branes, which are non-supersymmetric and uncharged. The
$p$-dimensional unstable D (D$p$) branes eventually decay into
stable D$(p-1)$ ones. This decay process is described by the
dynamics of a tachyon field $T$. The effective action of this
unstable brane system in low energy approximation is conjectured
to be of the Dirac-Born-Infeld (DBI) type form \cite{Garousi:2000tr,
Bergshoeff:2000dq,Sen:2002an}, as derived from string theory:
\begin{equation}
   S=-\int d^{p+1} x \sqrt{-g}V(T)\sqrt{1+
   \partial_\mu{T}\partial^\mu{T}},
\end{equation}
where the effective potential $V(T)$
takes its maximal value at $T=0$ and vanishes asymptotically as
$T$ tends to infinity.

The homogeneous decay of the tachyon field involves the field
rolling to its vacuum at $T=\pm\infty$, towards a state which can
be characterised as a pressureless fluid without propagating
modes. This is consistent with the notion that the open string
states disappear from the spectrum as the brane decays.

This decay can lead to a period of inflation
\cite{Mazumdar:2001mm,Gibbons:2002md}, although realistic models
are not easy to realise
\cite{Fairbairn:2002yp,Choudhury:2002xu,Kofman:2002rh,Sami:2002fs,
Shiu:2002qe}. One
reason is that the tachyon field perturbations around the unstable
vacuum must contain an unstable mode whose (negative) eigenvalue
is the tachyon mass squared, which is of order the string scale.
This leads to an $\eta$-problem in tachyonic inflation
\cite{Fairbairn:2002yp,Chingangbam:2004ng}.

In order to study the dynamics away from the unstable vacuum, at
least two supposed potential forms are often used:
$V(T)=V_m/\cosh{(\beta T)}$ (where $\beta=1$ for the bosonic
string and $\beta=1/\sqrt{2}$ for superstrings), and
$V(T)=\exp{(-T^2/2)}$. Both potentials lead to the correct value
of the mass of the tachyon on the D-branes. Here, we chose the
former because it contains as a static classical solution a
periodic array of static solitonic kink-antikink solutions, which
is known to exist as an exact solution in the deformed Conformal
Field Theory
\cite{Sen:1998tt,Sen:1998ex,Kutasov:2003er,Lambert:2003zr}. These
kinks have exactly the right tension to be understood as branes
and antibranes with one fewer dimension.

A similar effective action with a complex tachyon field describes
an unstable brane-antibrane system decaying to a network of stable
branes with two fewer dimensions, and in the case of D3 -
$\overline{\textrm{D}}3$ system, has been used to realise a
version of hybrid inflation terminating in the production of
cosmic superstrings \cite{Dvali:1998pa,Sarangi:2002yt,Choudhury:2003vr}.

In this paper we examine the low energy effective field theory
following from the tachyon effective action in the background of
static brane-antibrane pairs, neglecting the gravitational and
Ramond-Ramond fields. In the limit that the thickness of the
branes tends to zero, and the tachyon field between them
approaches the vacua at $T=\pm\infty$, we should expect
interactions between the branes to vanish and the resulting
spectrum should represent the effective field theory around
isolated branes.

The fluctuation spectrum in an effective action
$S=-\int d^{p+1} x \sqrt{-g} V(T) (1+
   \partial_\mu{T}\partial^\mu{T})$,
with the potential $V=\exp{(-T^2/2)}$ has been discussed in
\cite{Hashimoto:2002ct}, noting that the Regge slope $\alpha'=1$.
We find a number of similarities and differences. There are zero
modes associated with the translational degree of freedom of the
system, and also the arbitrary maximum value of $|T|$ between the
branes. We have numerical evidence that the lowest excited states
have a four-fold degeneracy in the limit that the brane thickness
vanishes. We find that the modes disappear from the bulk as the
field tends to the vacuum between the branes, consistent with the
absence of propagating modes of the tachyon field in the
homogeneous case. In the same limit, 4 degenerate zero-modes of
the tachyon field appear, two of which can be identified with the
transverse fluctuations of the branes.

Finally we study the dynamics of the zero mode corresponding to
changes in $T_0$, the maximum value of $|T|$ between the branes,
deriving the effective action for this modulus. We find a family
of slow-motion solutions corresponding to the field oscillating
between the unstable D$p$ brane and the stable D$(p-1)$,
$\overline{\textrm{D}}(p-1)$ pair.

This also has an interesting implication for the dynamical
production of D$(p-1)$-branes. In the unstable system, small
fluctuations drive the tachyon field $T(x)$ to roll from the top
of its potential $V(T)$.  D$(p-1)$-branes form as topological
defects (kinks or solitons) due to symmetry breaking
\cite{Shiu:2002xp,Cline:2002it,Sen:2003tm}. The time-dependent
tachyon field generically reaches a singularity in finite time,
either at the kinks \cite{Barnaby:2004dz,Cline:2003vc} or at
caustics between the kinks \cite{Felder:2002sv}. Our slow-motion
solution of the time-dependent modulus in the moduli space reaches
infinite gradient at the kink in finite time, consistent with
Ref.\ \cite{Barnaby:2004dz,Cline:2003vc}, but can be evolved
through the singularity.

The paper is constructed as follows. We first discuss the static
solutions of the tachyon equation of motion derived from the DBI
effective action in Sec.\ 1. Based on the solution, fluctuation
modes both in approximate analysis and in general form are
discussed in Sec.\ 2. In Sec.\ 3, we further investigate the
motion in the moduli space. Sec.\ 4 gives conclusions of the
paper.

\section{Static solutions to the field equations}
\label{sec:sta-sol}

The equation of motion derived from the DBI action is:
\begin{equation}
   \Box T+(g^{\mu\nu}\partial{T}\cdot\partial{T}-
   \partial^\mu{T}\partial^\nu{T})\partial_\mu\partial_\nu{T}
-\frac{V'}{V}(1+\partial{T}\cdot\partial{T})=0.
\end{equation}

We are interested in time-independent solutions, neglecting the
gravitational and Ramond-Ramond fields of the brane. It is known
that there are solutions depending on only one space coordinate,
which we denote $x$ \cite{Hashimoto:2002ct,Kim:2003in,Sen:2004nf}.
In this case, the static but inhomogeneous field equation reduces
to:
\begin{equation}
   \left(\partial_x{T}\right)^2=\left(\frac{V}{V_0}\right)^2-1,
\end{equation}
where $V_0=V(T_0)$ is the minimum value of the potential, which
occurs at points $x=x_0$ satisfying $\partial_x T(x)|_{x_0}=0$.
$V_0$ is not zero providing $T$ remains finite. The solution is:
\begin{equation}
\label{e:Tsol}
   T(x)=\frac{1}{\beta}\mathrm{arcsinh}\left[\sin(\beta (x-x_m))\sqrt{\left
   (\frac{V_m}{V_0}\right)^2-1} \right],
\end{equation}
which is a regular array of kinks and anti-kinks, with period
$2\pi/\beta$, located at $x=2n\pi/\beta+x_m$ and
$x=(2n-1)\pi/\beta+x_m$ ($n$ is integer) respectively. $x_m$ is
set to be the positions where the maximum potential $V_m$ is
located. The energy density is peaked near the zeros of $T$, where
$V=V_m$, and in the limit that $V_0\to 0$ the peaks become
infinitely thin, and we identify the kinks with daughter branes
and anti-branes with one dimension lower. The minimum potential
$V=V_0$ and the maximum amplitude of the tachyon field
$|T_0|=\mathrm{arccosh}(V_m/V_0)/\beta$ is reached at
$x_0=(n+\frac12)\pi/\beta+x_m$, which is the closest approach to
the vacuum of the system.

The energy of this system is
\begin{equation}
\label{e:En}
   E=\int d^px V(T) \sqrt{1+(\partial_x{T})^2}=\frac{1}{V_0} \int d^{p-1}x \int dx V^2.
\end{equation}
The energy per unit $(p-1)$-dimensional volume
in one period of the tachyon field is:
\begin{equation}
 \sigma_{1}=\frac{\pi V_m}{\beta}
\end{equation}
with the specific potential adopted in this paper. The tension of
the parent D$p$-brane is $\T_p = V_m$, so this is the correct
tension for a D$p-1$-brane, $\T_{p-1}$.

\section{Fluctuation modes}
\label{sec:flu}

We study the fluctuation modes by linearizing the variation,
$T(t,x)\rightarrow T+\delta{T}(x)e^{-i\om t}$. Following
\cite{Brax:2003rs}'s procedure, which makes the substitution
$dx=({V_0}/{V})dz$ and $\delta{T}=({V}/{V_0})\phi$, we have
\begin{equation}
  \left( -\frac{d^2}{dz^2}+U(z)\right)\phi=\omega^2\phi,
\end{equation}
where
\begin{equation}
  U(z)=\frac{1}{\phi_0}\frac{d^2\phi_0}{dz^2}.
\end{equation}
Here, $\phi_0 \propto({dT}/{dx})\sqrt{{V_0}/{V}}$ is the zero mode
($\omega^2=0$). We can get different eigenstates with different
boundary conditions and their corresponding eigenvalues $\omega^2$
from this equation.

More precisely, we can get the simple form of $U(z)$:
\begin{equation}
  U(z)=\frac{\beta^2}{4}\left\{1+\left(\frac{V_0}{V_m}\right)^2-3
  \left[\left(\frac{V(z)}{V_m}\right)^2+\left(\frac{V_0}{V(z)}\right)^2
  \right]\right\}.
\end{equation}
The maximum value of $U(z)$ is at $V(z)=\sqrt{V_0}$, where
$U=\beta^2/4+O((V_0/V_m)^2)$. The minimum potential
$U(z)=-\beta^2/2+O((V_0/V_m)^2)$ happens at both $V(z)=V_m$ and
$V(z)=V_0$. With the fluctuation equation, we can discuss the
fluctuation modes and the discrete spectrum $\omega^2$.

\subsection{Approximate fluctuation modes near kinks}

We start by studying the behaviour near the kink or brane, where
$T(z) \simeq 0$ and $V(z)\simeq V_m$ as mentioned above. We will
assume that the kinks are well separated in comparison to their
width, which implies that $V_0/V_m \ll 1$. In the limit
$V(z)\rightarrow V_m$, ${dT}/{dz}\simeq \pm 1$ since
${dT(z)}/{dz}=({dT}/{dx})({dx}/{dz})=\pm \sqrt{1-({V_0}/{V})^2}$.
Hence, near a kink, $T(z)\simeq \pm (z+c)$ and $V(z)\simeq
V_m/\cosh{(\beta (z+c))}$, where $c$ is a constant. Therefore, we
have,
\begin{equation}
  \frac{d^2\phi}{dz^2}+\left[\omega^2+\frac{\beta^2}{4}
  \left(\frac{3}{\cosh^2{(\beta (z+c))}}-1\right)\right]\phi=0,
\end{equation}
which is a well explored equation. Following standard
procedures \cite{LanLif} we can recast the
equation by setting $\xi=\tanh{(\beta (z+c))}$:
\begin{equation}
  \frac{d}{d\xi}(1-\xi^2)\frac{d}{d\xi}\phi+\left[\frac{3}{4}
  -\left(\frac{1}{4}-\frac{\omega^2}
  {\beta^2}\right)\frac{1}{1-\xi^2}\right]\phi=0,
\end{equation}
which has solutions
\begin{equation}
  \phi=(1-\xi^2)^{\epsilon/2}F[\epsilon-\tfrac{1}{2},\epsilon
  +\tfrac{3}{2},\epsilon+1,\tfrac{1}{2}(1-\xi)],
\end{equation}
where $\epsilon=\sqrt{\frac{1}{4}-\frac{\omega^2}{\beta^2}}$.

If $\phi$ is finite, we have that $m=\frac{1}{2}-\epsilon$ must be
a non-negative integer. The condition $\epsilon \geq 0$ or
$\omega^2<\beta^2/4$ gives that $m$ can only be zero, i.e.,
$\omega^2=(1-m)m\beta^2=0$. Thus there is only a zero fluctuation
mode with $\om^2$ below $\beta^2/4$, the maximum of the potential
$U(z)$. This mode is of course just the zero mode associated with
the translational parameter of the solutions. Modes with
$\omega^2<\beta^2/4$ are bound states, confined to the kink, and
so we conclude that there are no bound states other than the zero
mode, at least in the thin kink limit.

\subsection{General fluctuation modes}

To obtain the generalized solutions, we need to know the form of
the $z$-dependent potential $V(z)$. From $dz=({V(x)}/{V_0})dx$, we
find
\begin{equation}
  \int dz=-\int \frac{1}{\beta V_0}\frac{dV}{\sqrt{(\frac{V}
  {V_0})^2-1}\sqrt{1-(\frac{V}{V_m})^2}}.
\end{equation}
With the reparametrization $t=V/V_0\in [1,t_m=V_m/V_0]$, it leads
to:
\begin{equation}
  \int_{z_0}^{z} dz'=-\int_{1}^{t}
  \frac{i}{\beta}\frac{1}{\sqrt{1-t'^2}\sqrt{1-k^2t'^2}}dt'.
\end{equation}
The integral can be evaluated in terms of an elliptic function of
the first kind:
$F(\varphi,k)=\int_{0}^{\sin{\varphi}} dx(1-x^{2})^{-\half}(1-k^{2}x^{2})^{-\half}$,
where the parameter $k=V_0/V_m$. With $t=\sin{\varphi}$, we have
$\beta (z-z_0)=\beta \Delta z=-i[F(\varphi,k)-K(k)]$, or we can write:
\begin{equation}
  F(\varphi,k)=i\beta \Delta z+K(k),
\end{equation}
where $K(k)$ is the complete elliptic function of the first kind:
$K(k)=F(\frac{\pi}{2},k)$. Its inverse is one of the Jacobi
functions,
\begin{equation}
  t=\frac{V(z)}{V_0}=\mathrm{sn}\left(i\beta \Delta z+K(k),k\right).
\end{equation}
Using the identities of the Jacobi functions \cite{GraRyz}
\begin{equation}
\sn(i\beta \Delta z+K(k),k)=\frac{\cn(i\beta \Delta z,k)}{\dn(i\beta \Delta
 z,k)}=\frac{1}{\dn(\beta \Delta z,k')}.
\end{equation}
where $k'=\sqrt{1-k^2}$, we can bring $U(z)$ into the form
\begin{equation}
  U(z)=\frac{\beta^2}{4}\left\{2-k'^2-3\left[\frac{k^2}{\dn^2(\beta
  \Delta z, k')}+\dn^2(\beta \Delta z, k')\right]\right\}.
\end{equation}
Notice that, in the following numerical calculations, we just
adopt $z$ instead of $\Delta z$ by setting $z_0=0$.

The curves of $T(z)$, $V(z)$ and $U(z)$ with $k=0.01$ are depicted
in Fig 1. They are all periodic, with periods $4K(k)/\beta$,
$2K(k)/\beta$ and $K(k)/\beta$ respectively, which comes from $\dn(u)$
being an even function with period $z=2K(k)/\beta$.

Since $K$ increases with $k$, the separations between two
neighboring branes becomes larger as $k$ decreases when taking $z$
as the coordinate, while remaining constant in the $x$-coordinate.
The daughter stable D$p-1$ branes and anti-D$p-1$ branes are
located at $z=(4n+1)K/\beta$ and $z=(4n-1)K/\beta$ respectively,
where the tachyon field $T(z)$ vanishes and the potential $V(z)$
peaks. The closest approach to the vacuum occurs at $z=2nK/\beta$,
where $V(z)$ gets its minimum value $kV_m$. Note that the plots
present only one period of the solution
$[-2K(k)/\beta,2K(k)/\beta]$.

From the figure, we can see that $T(z)$ is indeed linearly
dependent on $z$ for most part of the numerical curve, especially
near the kink positions. The smaller the value of $k$, the larger
the proportion of the solution occupied by the linear section,
confirming the validity of the approximation adopted in the
previous subsection.

\begin{figure}
\centering
\includegraphics[angle=0,scale=0.4]{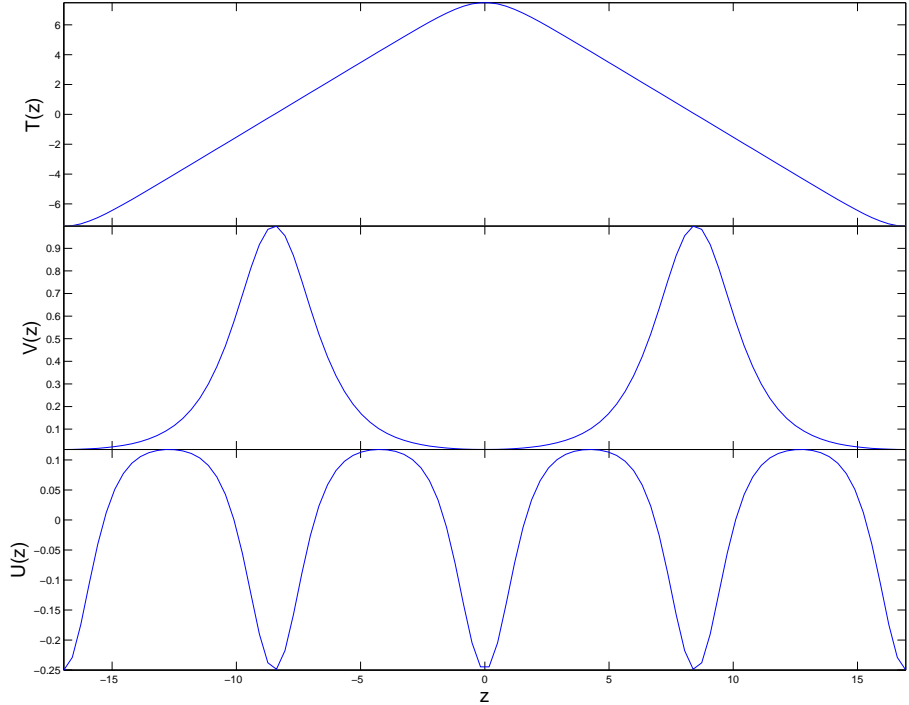}
\caption{\label{f:f1} Plots of $T(z)$, $V(z)$ and
$U(z)$ within a period of the tachyon field $[-2K(k)/\beta,2K(k)/\beta]$.
Here $k=0.01$.}
\end{figure}

\subsection{Numerical solution of eigenvalue equations}

The periodicity of the solution and the potential for the
fluctuations means that we need consider the range $0\le z \le
K/\beta$, with eigenfunction taking either Dirichlet (D) or von
Neumann (N) boundary conditions at each boundary. Hence there are
four boundary conditions (BCs) which are listed and labelled in
Table \ref{t:t1}.

We solve the eigenvalue equations by numerical integration,
``shooting" for the correct boundary condition at $z = K/\beta$,
using the 4th/5th order Runge-Kutte integrator built into MATLAB,
\texttt{ode45}. We adopted an absolute tolerance of $10^{-9}$ and
a relative tolerance of $10^{-6}$ for all calculations in this
paper.

\begin{table}
\begin{center}
\begin{tabular}{|c|c|c|c|}
\hline
BC(1): ND &BC(2): NN & BC(3): DD &BC(4): DN\\
\hline
$\phi'(0)=0$&$\phi'(0)=0$&$\phi(0)=0$&$\phi(0)=0$\\
$\phi(K/\beta)=0$&$\phi'(K/\beta)=0$&$\phi(K/\beta)=0$&$\phi'(K/\beta)=0$\\
\hline
$\phi(0)=1$&$\phi(0)=1$&$\phi'(0)=1$&$\phi'(0)=1$\\
\hline
\end{tabular}
\caption{\label{t:t1} The four possible boundary conditions for
the eigenfunctions of small fluctuations in the tachyon field
about the static kink-antikink array. The letters D and N denote
Dirichlet and von Neumann boundary conditions respectively. The
last row gives the initial conditions for the numerical solution
of the eigenvalue equation.}
\end{center}
\end{table}

\subsubsection{Fluctuation modes of low lying states}

\begin{figure}
\centering
\includegraphics[angle=0,scale=0.27]{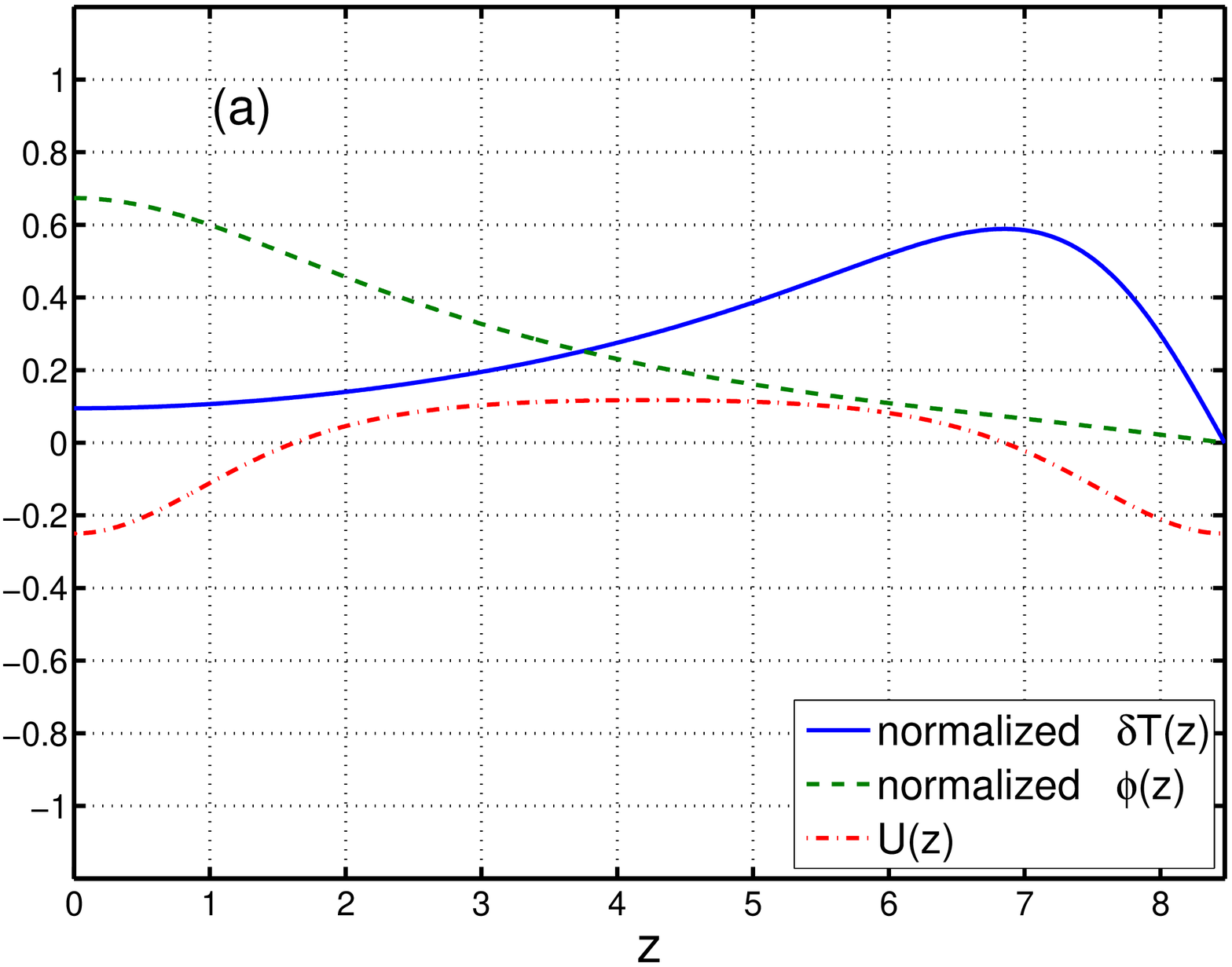}
\includegraphics[angle=0,scale=0.27]{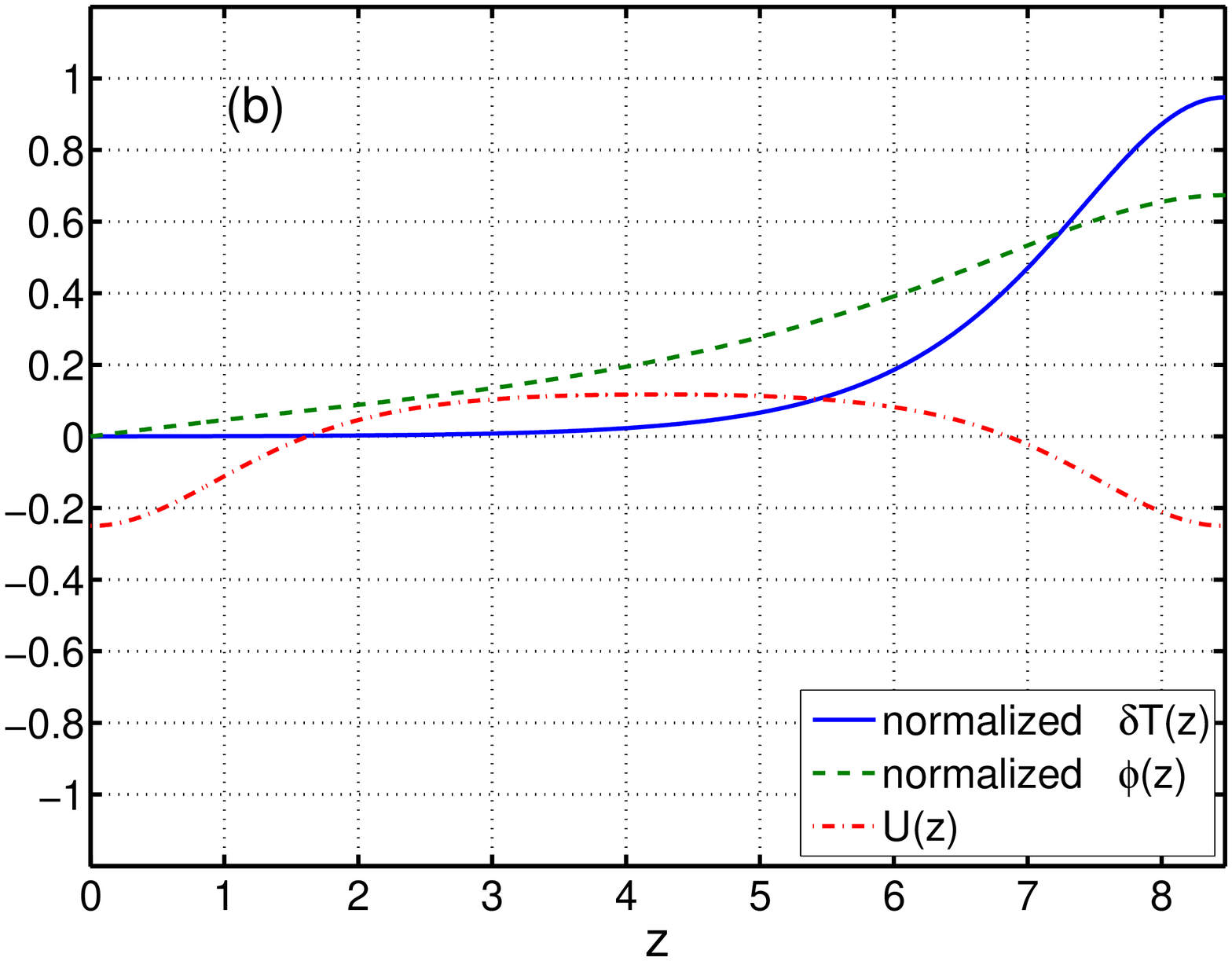}
\includegraphics[angle=0,scale=0.27]{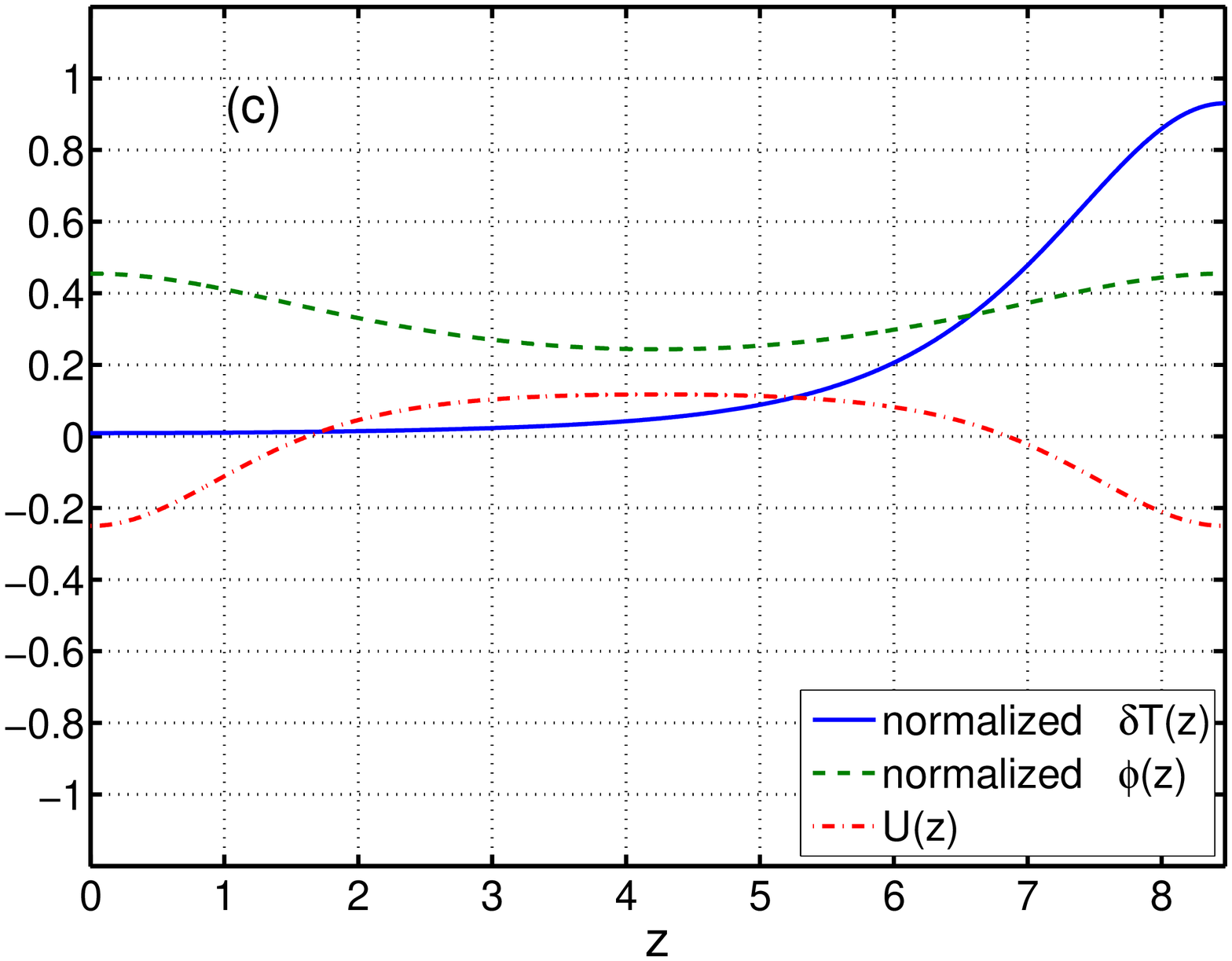}
\includegraphics[angle=0,scale=0.27]{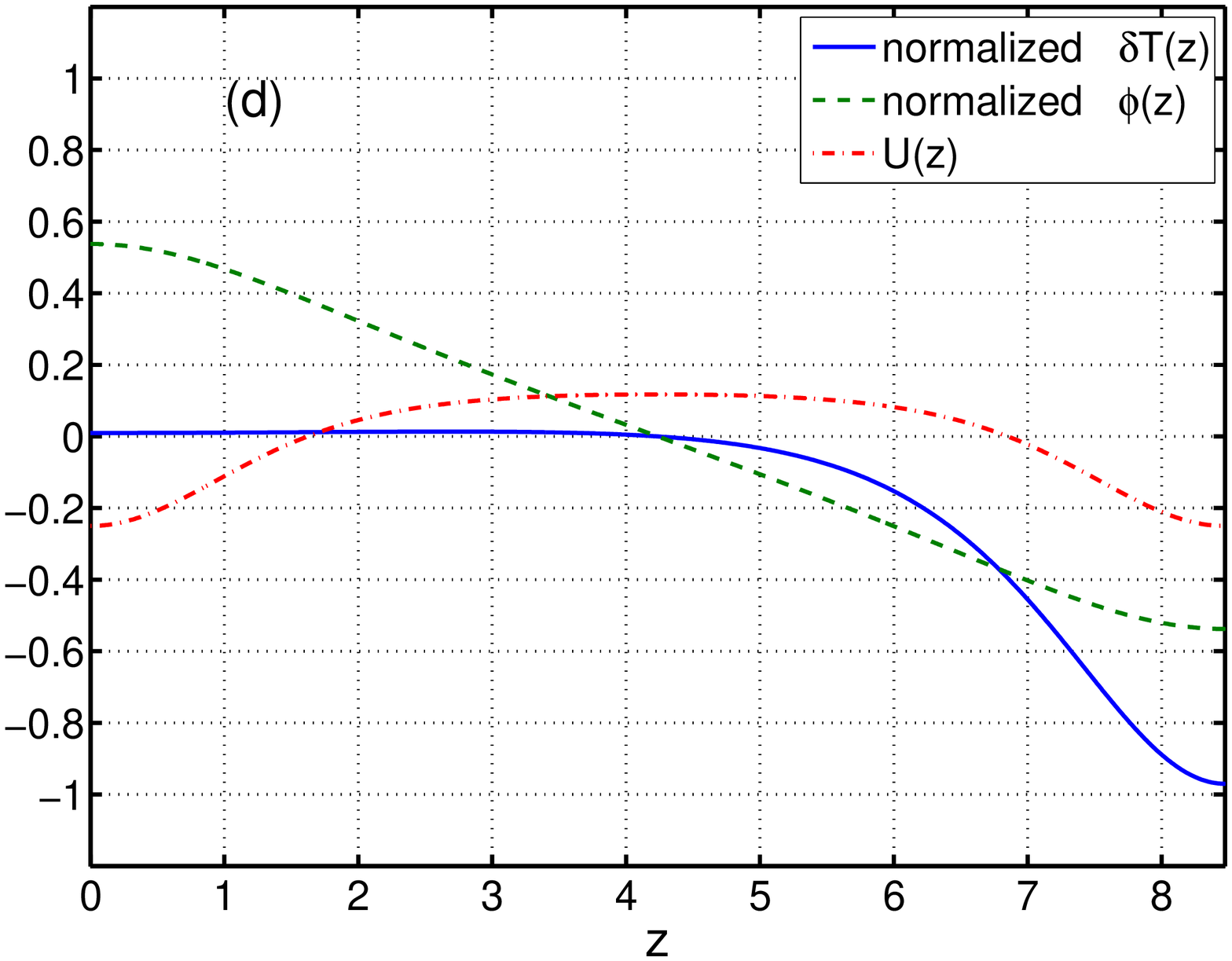}
\caption{\label{f:f2} Normalized numerical solutions, $\delta
\tilde{T}(z)$ and $\tilde{\phi}(z)$, of lowest eigenvalue modes
with $k=0.01$, plotted with the potential $U(z)$ for comparison.
(a) BC(1) and (b) BC(4) have eigenvalues which are consistent with
zero, respectively $\omega^2=-1.0672\times10^{-8}$ and
$\omega^2=-2.0095\times10^{-9}$.
The other two figures are both under BC(2) and have two bound
states with (c) $\omega^2=-0.0270$ (which we call $0^{-}$) and (d)
$ \omega^2=0.0433 $ (which we call $0^{+}$). BC(3) (not shown in
the figure) has no eigenvalues close to zero.}
\end{figure}

\begin{figure}
\centering
\includegraphics[angle=0,scale=0.27]{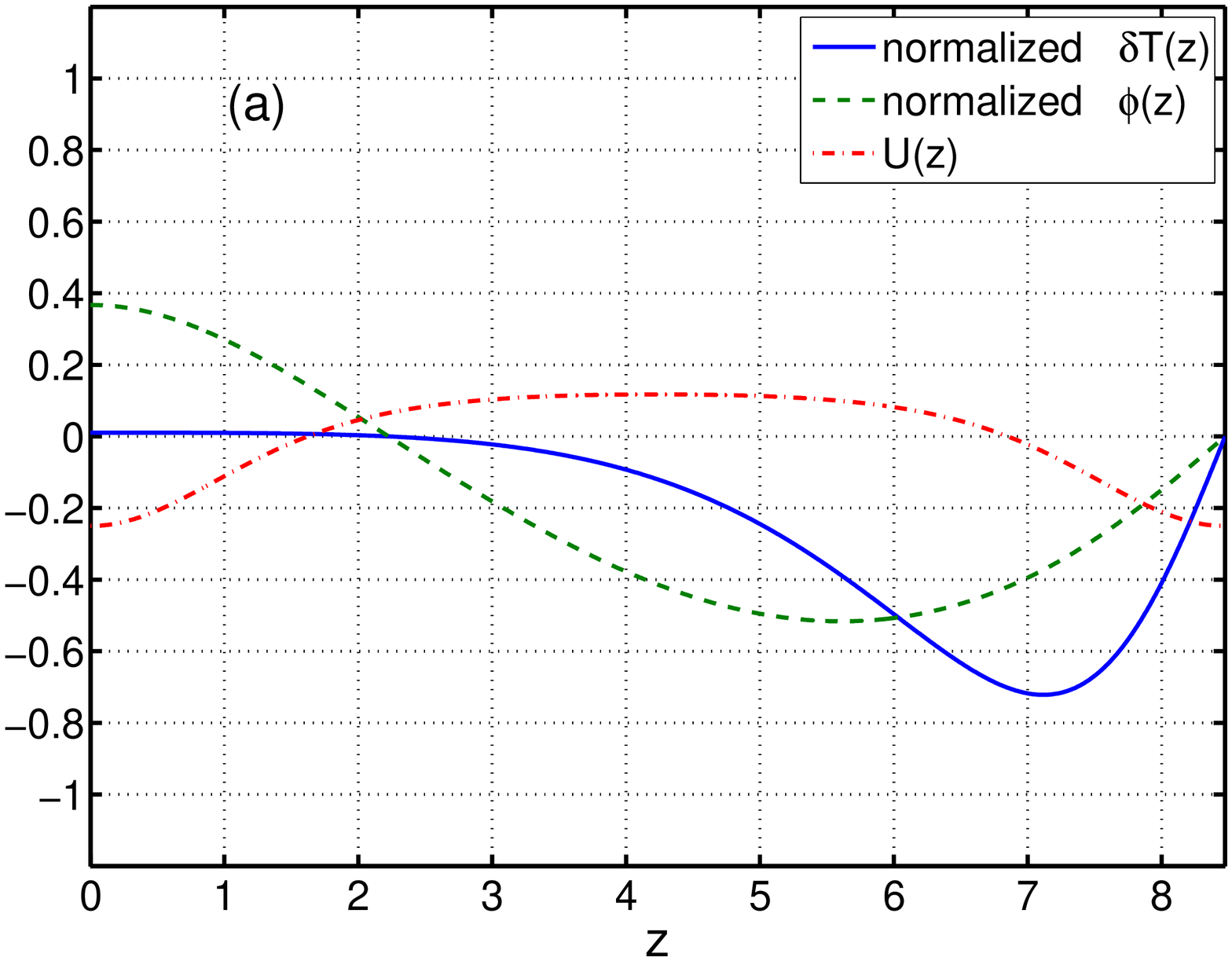}
\includegraphics[angle=0,scale=0.27]{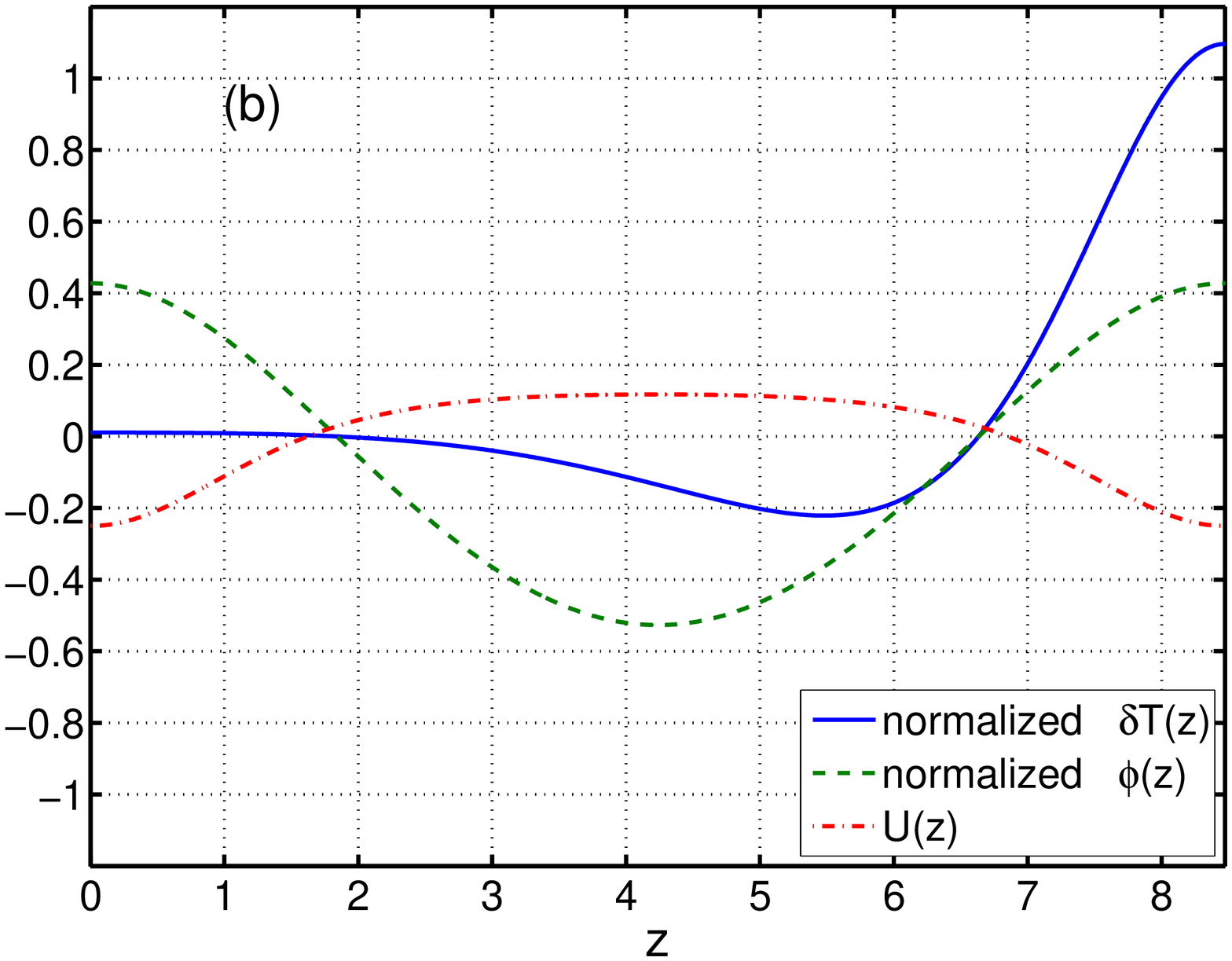}
\includegraphics[angle=0,scale=0.27]{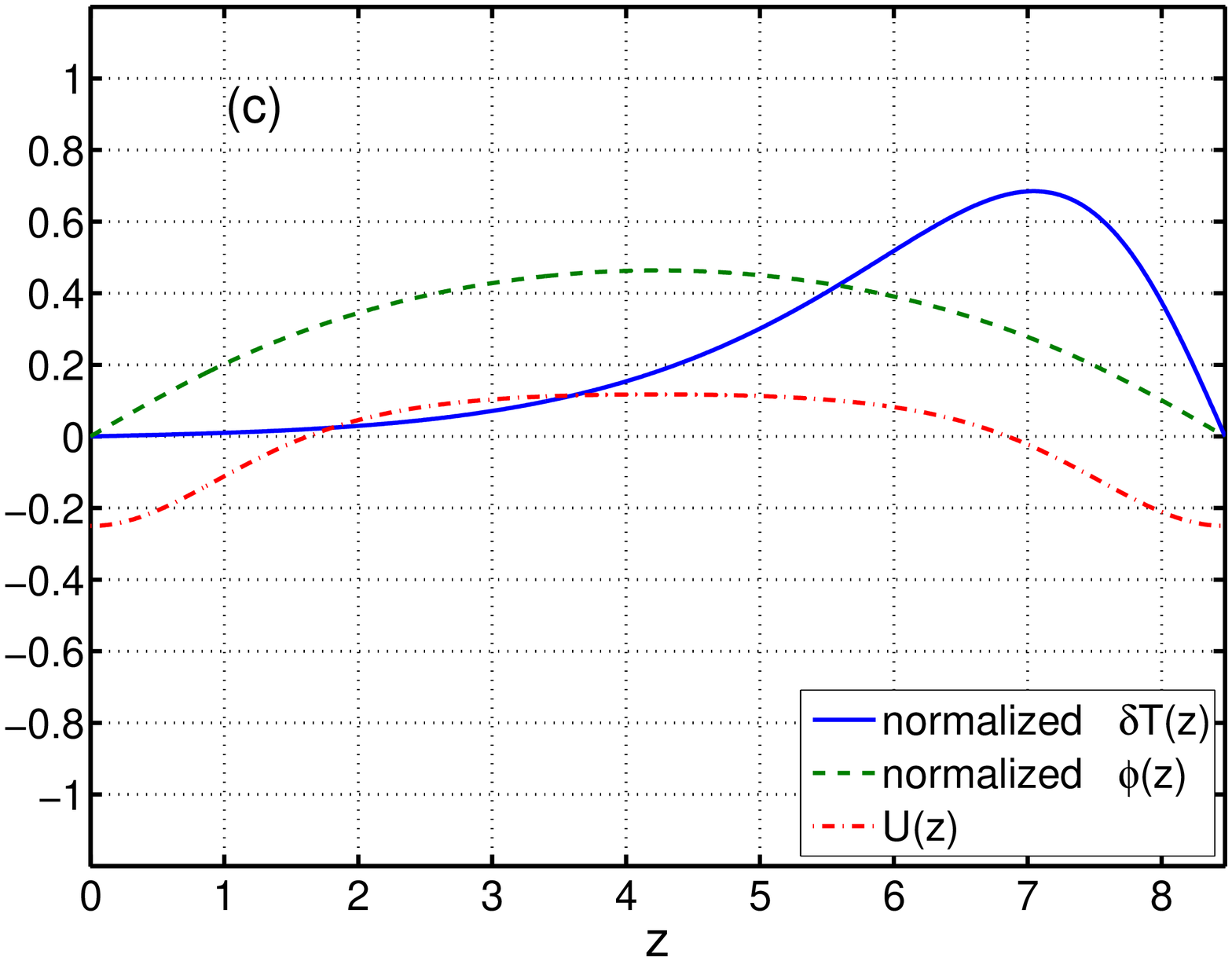}
\includegraphics[angle=0,scale=0.27]{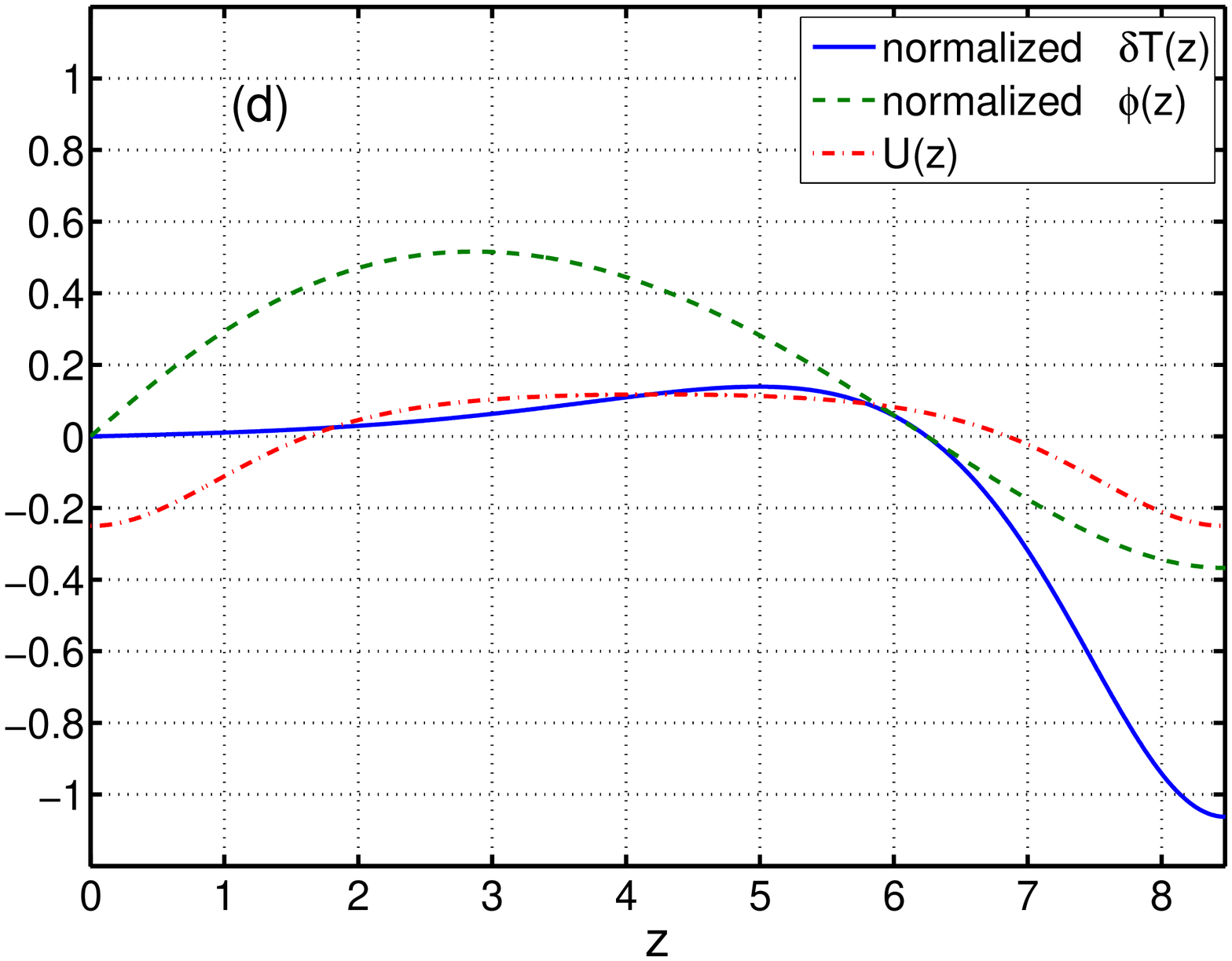}
\caption{\label{f:f3} Normalized numerical solutions, $\delta
\tilde{T}(z)$ and $\tilde{\phi}(z)$, of the first excited states
with $k=0.01$ under the four boundary conditions: (a) BC(1), (b)
BC(2), (c) BC(3) and (d) BC(4). Note that BC(1) and BC(4) have the
same eigenvalue.}
\end{figure}

\begin{figure}
\centering
\includegraphics[angle=0,scale=0.27]{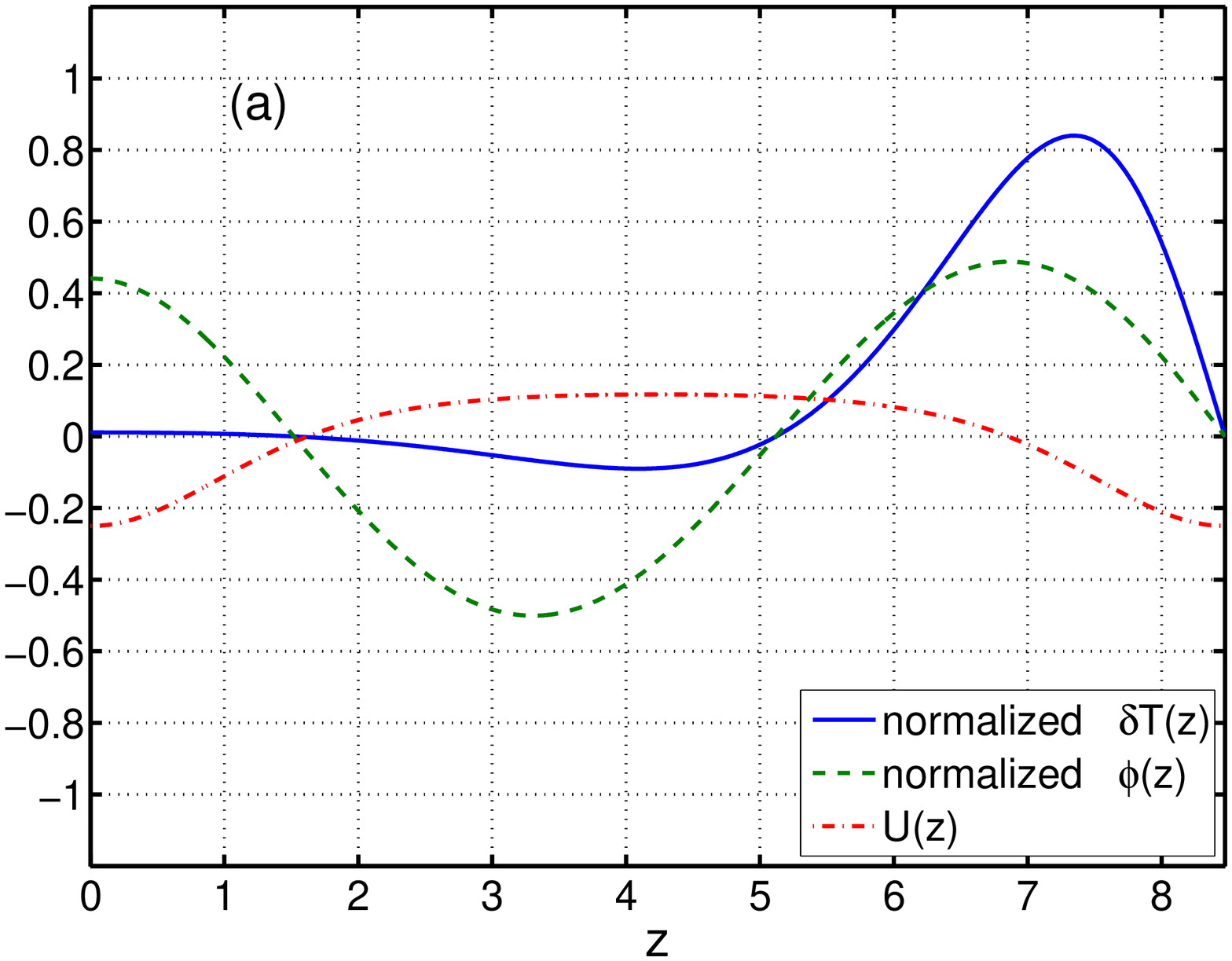}
\includegraphics[angle=0,scale=0.27]{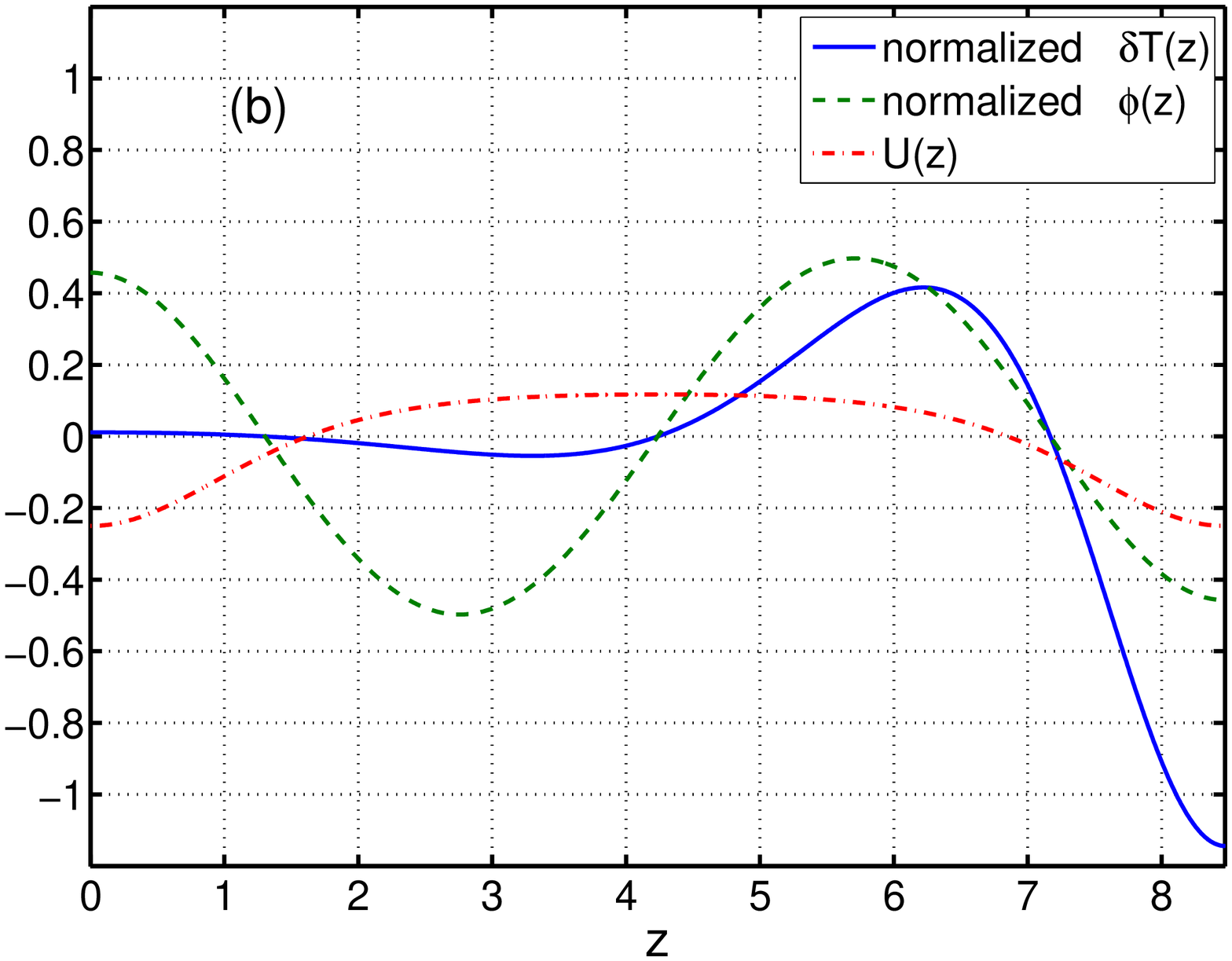}
\includegraphics[angle=0,scale=0.27]{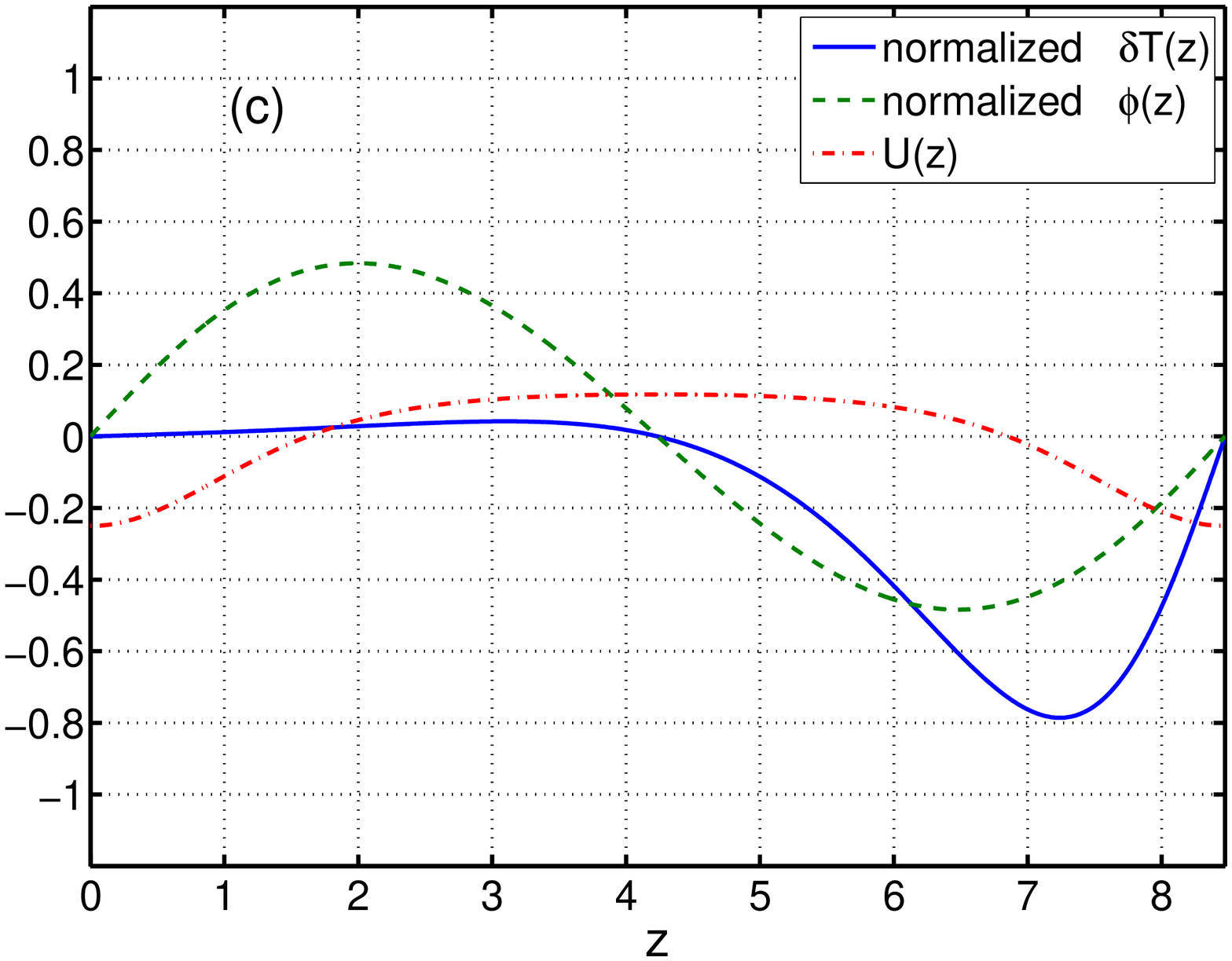}
\includegraphics[angle=0,scale=0.27]{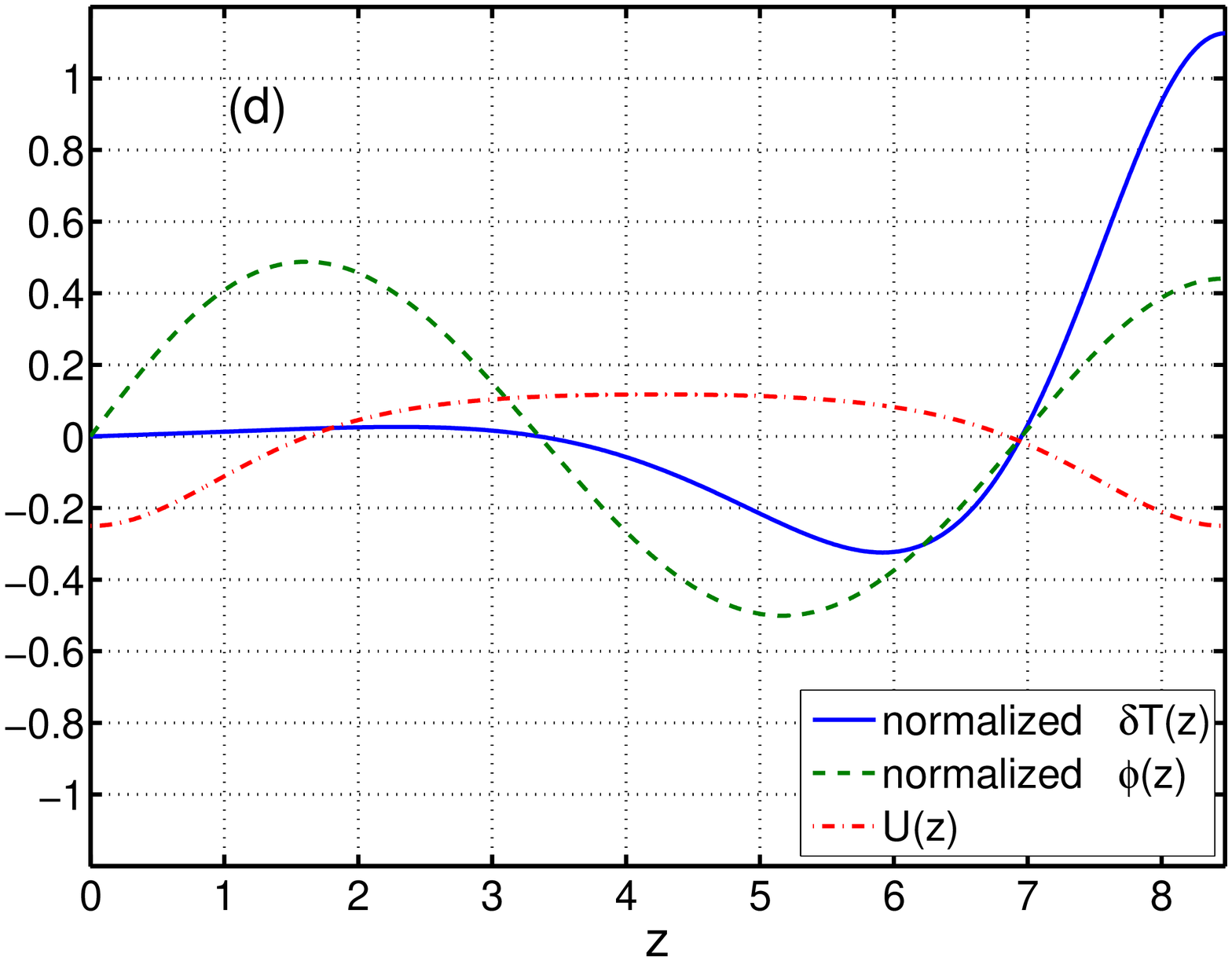}
\caption{\label{f:f4} Normalized numerical solutions, $\delta
\tilde{T}(z)$ and $\tilde{\phi}(z)$, of second excited states with
$k=0.01$ under the four boundary conditions: (a) BC(1), (b) BC(2),
(c) BC(3) and (d) BC(4). BC(1) and BC(4) have the same
eigenvalue.}
\end{figure}

We first set $k=0.01$. In Figs \ref{f:f2}, \ref{f:f3} and
\ref{f:f4}, normalized numerical solutions are presented in groups
of four, whose eigenvalues are close together. We refer to them as
zero mode states, first excited states and second excited states
respectively. We denote normalised solutions by a tilde, and plot
both $\tilde\phi(z)$ and $\delta \tilde{T}(z)$, which are defined
as
\begin{equation}
\delta \tilde{T}(z) = \frac{\delta
T(z)}{\int_{0}^{K/\beta}{|\delta T(z^{\prime})|^2 dz^{\prime}}}
\end{equation}
and
\begin{equation}
\tilde{\phi}(z) =
\frac{\phi(z)}{\int_{0}^{K/\beta}{|\phi(z^{\prime})|^2
dz^{\prime}}}.
\end{equation}

Similarities and symmetries are clearer when looking at
${\phi}(z)$, as $\delta{T}(z)$ has a factor $V(z)/V_0$ which does
not share the periodicity of the eigenfunctions, and suppresses
the function away from the kinks. For example, eigenfunctions with
BC(1) and BC(4) are related by the symmetry $z \to K/2\beta-z$,
and therefore the eigenvalues are the same.

Because of the factor $V(z)/V_m$,  the fluctuations away from the
kinks are greatly suppressed when plotted in terms of $\delta T$,
and the eigenfunctions look very similar near the (anti)kink at $z
= K(k)/\beta$.

Fig.\ \ref{f:f1} contains the four states whose eigenvalues are
very close to zero. Indeed, two with BC(1) and BC(4), and
eigenvalues $\omega^2=-1.0672\times10^{-8}$ and
$\omega^2=-2.0095\times10^{-9}$, are consistent with being
numerical approximations to the true zero modes corresponding to
the degeneracies in position $x_0$ (Fig.\ \ref{f:f1}a) and
amplitude $T_0$ (Fig.\ \ref{f:f1}b) of the solution. Not only do
they have very small eigenvalues, they also have the correct
number and position of nodes.

We also have two low-lying modes with BC(2), one with a small
negative eigenvalue, which we call $0^-$, and the other with a
small positive one $0^+$. The negative eigenvalue reflects an
instability in the system towards annihilation, where the
kink-antikink pair moves together. There is no low-lying state
with BC(3).

We can understand at least the ordering of the eigenvalues by
studying the number of nodes of the eigenfunctions, using the
standard result that more nodes means a higher eigenvalue.

We note first that the background has periodicity $4K(k)$, in
which the eigenfunctions must also be periodic, and therefore have
an even number of nodes. The lowest eigenvalue is associated with
BC(2), which allows an eigenfunction with no nodes. BC(1,4) allow
only 2 nodes, which are the zero modes. BC(2) also allows the
4-node eigenfunction shown in Fig.\ \ref{f:f2}(d). BC(3) permits a
minimum of 4 nodes which, because of its Dirichlet boundary
conditions, are forced to be at the minimum of the potential
$U(z)$, and it must therefore have a higher eigenvalue than the
4-node BC(2) eigenfunction, which peaks where the potential is
lowest.

\begin{figure}
\centering
\includegraphics[angle=0,scale=0.27]{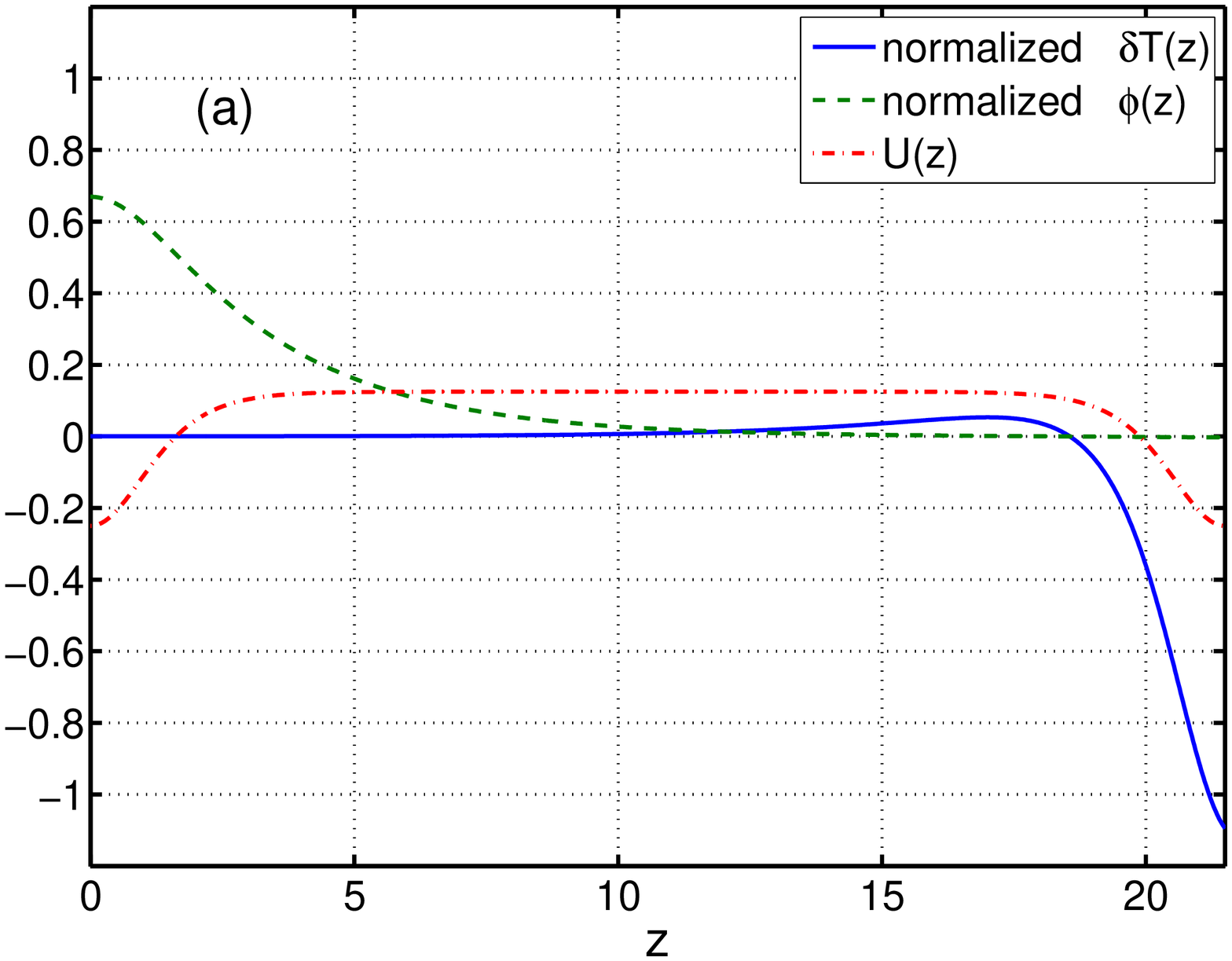}
\includegraphics[angle=0,scale=0.27]{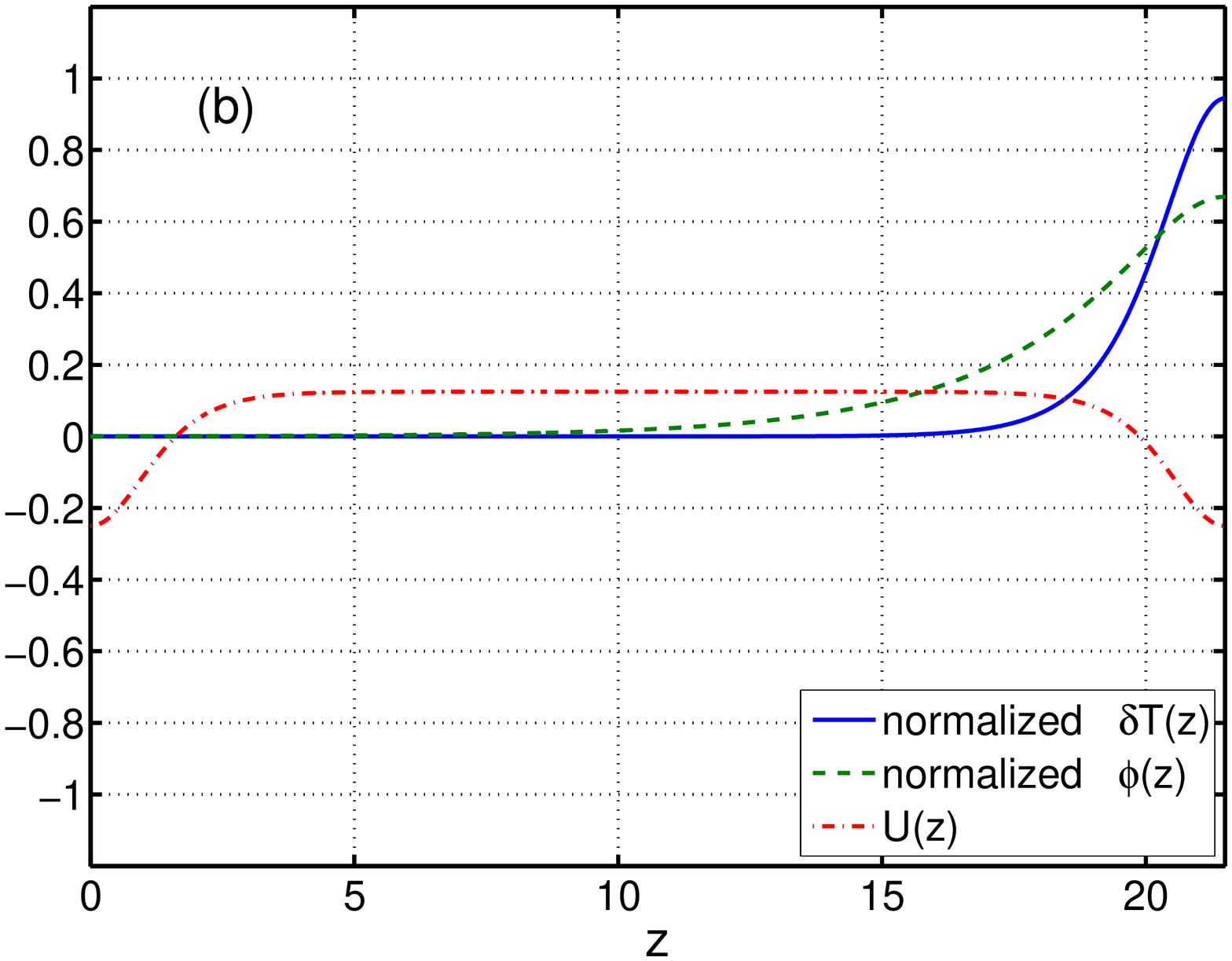}
\includegraphics[angle=0,scale=0.27]{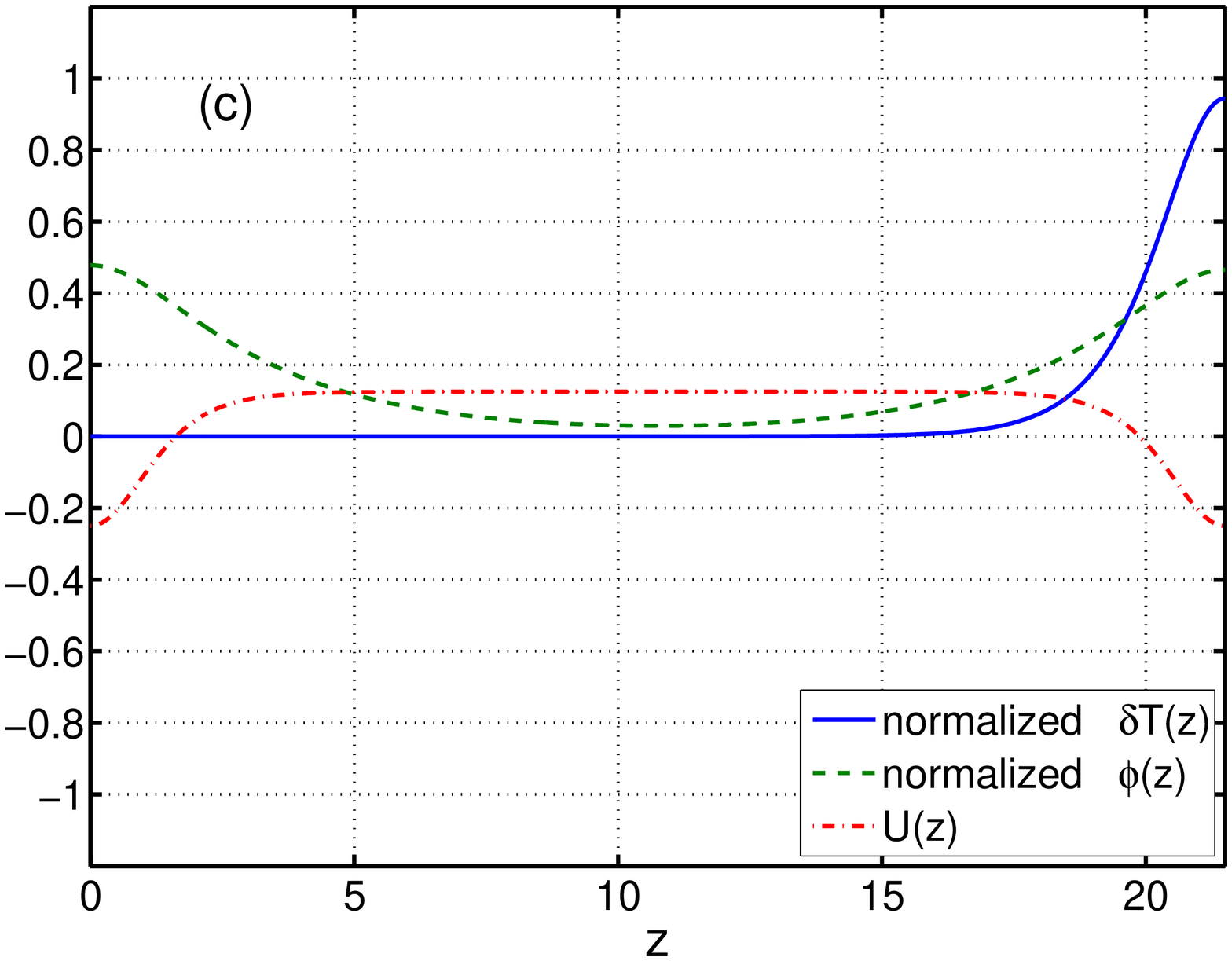}
\includegraphics[angle=0,scale=0.27]{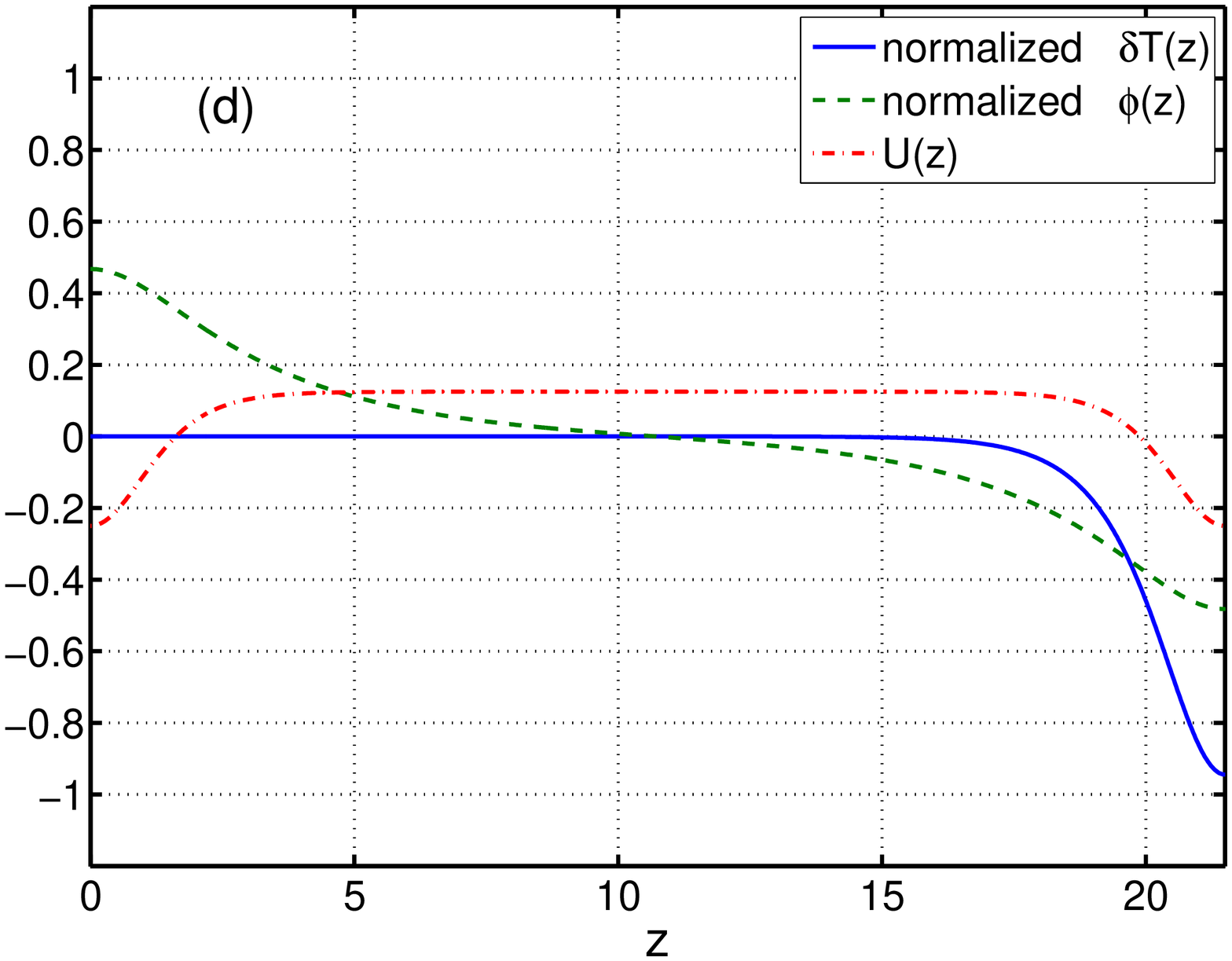}
\caption{\label{f:f5} Normalized numerical solutions, $\delta
\tilde{T}(z)$ and $\tilde{\phi}(z)$, of zero modes with
$k=10^{-6}$. As for the case $k=10^{-2}$, (a) BC(1)
and (b) BC(4) have eigenvalues very close to zero, which are
respectively $\omega^2=-1.0765\times10^{-8}$ and
$\omega^2=1.6806\times10^{-5}$. BC(2) also has two bound states
with (c) $0^{-}$ mode: $\omega^2=-3.0842\times10^{-4}$ and (d) $0^{+}$
mode: $\omega^2=3.6063\times10^{-5}$.
BC(3) has no zero eigenvalue.}
\end{figure}

\subsubsection{Fluctuations with smaller $k$}

Stable BPS daughter D$(p-1)$ branes form as $V_0$ approaches zero
at the end of the tachyon condensation, so we should check the
fluctuation behaviour as $k \rightarrow 0$. However, we were
unable to get numerical solutions with very small $k$. In Fig.\
\ref{f:f5}, we show the zero mode fluctuation with $k=10^{-6}$.
Here it is even more obvious that fluctuations have support mostly
near the brane position, both in $\phi$ and $\delta T$. The higher
modes have similar properties. Therefore, we can expect that the
fluctuations are confined to the branes more and more strongly
when $k$ or $V_0$ tend to zero. We can interpret this as being
consistent with the observation \cite{Sen:2004nf} in the
homogeneous case that there are no propagating tachyon modes
($\omega^2 \rightarrow 0$) in the vacuum when $V_0 \rightarrow 0$.

\subsubsection{Eigenvalues with varying $k$}

\begin{figure}
\centering
\includegraphics[angle=-90,scale=0.24]{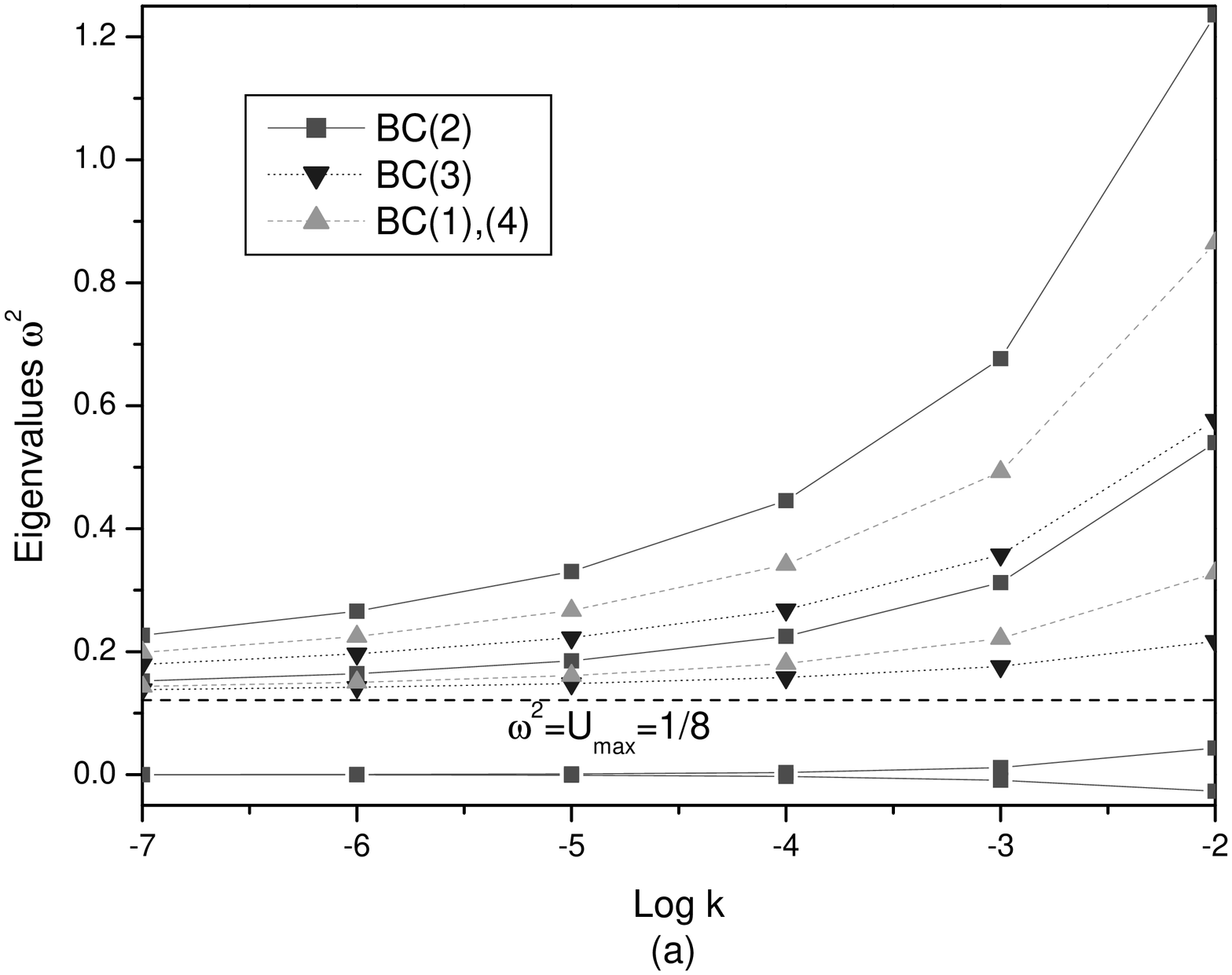}
\includegraphics[angle=-90,scale=0.24]{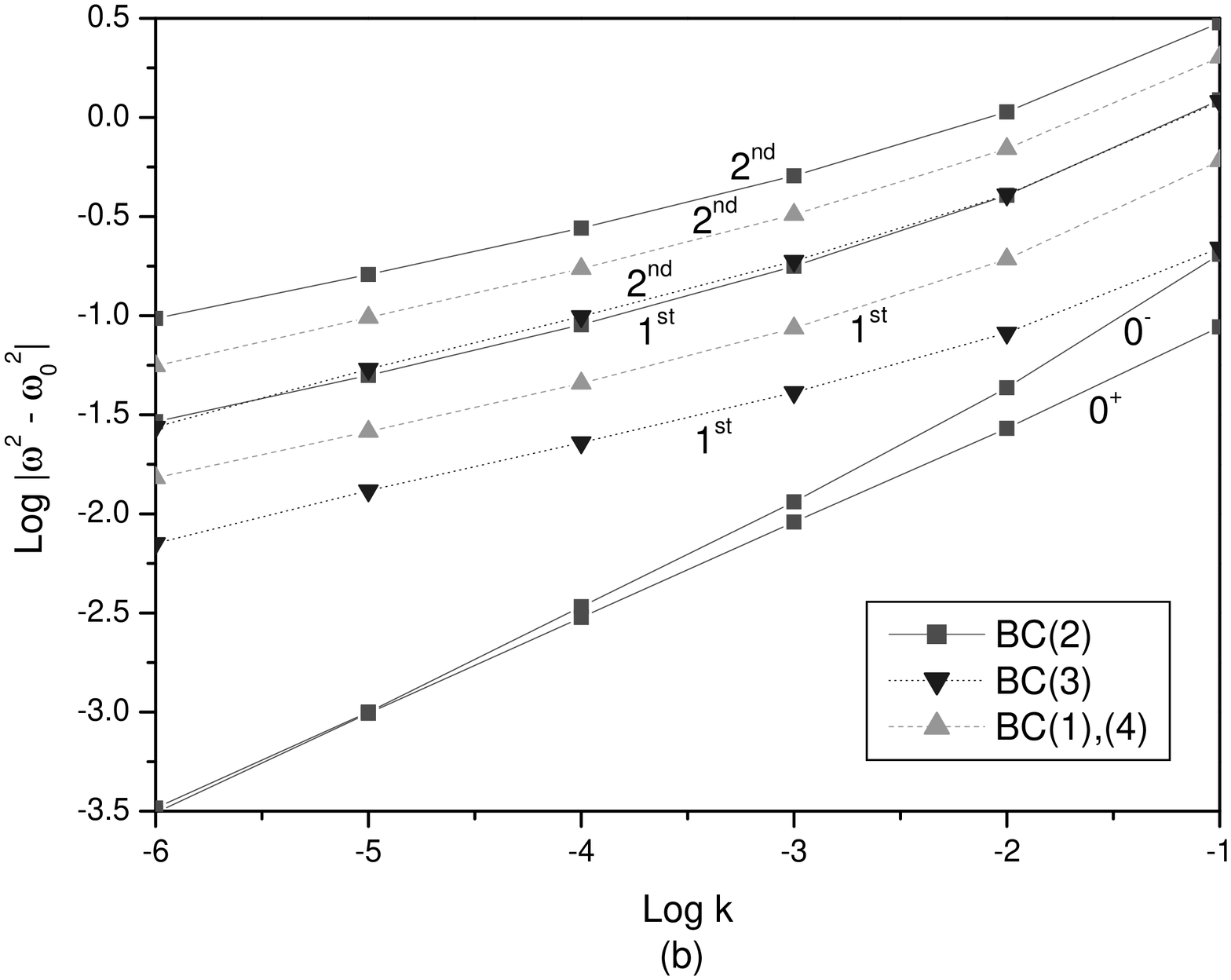}
\caption{\label{f:f6} Eigenvalues of zero modes, first excited and
second excited states against $\log k$ corresponding to the four
BC respectively. Note that BC(1) and BC(4) have the same
eigenvalues for all states because of symmetry.}
\end{figure}

We also investigate the eigenvalues against $\log k$ for all
boundary conditions in Fig.\ \ref{f:f6}, down to $\log k = -7$. We
could not get eigenvalues for smaller $k$ for reasons of numerical
precision.

As shown in Fig.\ \ref{f:f6}a, eigenvalues $\omega^2$ seem to
approach constant values with $k$ decreasing (which corresponds to
letting $V_0$ tend to $0$ while keeping $V_{max}=1$ constant). We
also see that the eigenvalues seem to assemble themselves into
groups of four, which we referred to earlier as zero modes, first
excited and second excited states. The grouping is not so clear
for the second excited states, but supporting evidence is shown in
Fig.\ \ref{f:f6}b, where we see an approximate power law
dependence for $\omega^2-\omega_{(n)\infty}^2$ against $k$,  where
$\omega_{(0)\infty}^2=0$ for the zero mode, $\omega_{(1)\infty}^2=
0.135$ for the first excited state and
$\omega_{(2)\infty}^2=0.169$ for the second one. Therefore, they
should be the degenerate eigenvalues of all the four BCs in the
limit of $T_0 \rightarrow \infty$ or $k \rightarrow 0$.

The only bound states (whose eigenvalues are between 0 and
$U_{max}\simeq 1/8$) are those associated with translations and
changes in $V_0$, corroborating our analysis of the previous
section with the approximate solution $T(z) \propto z$.

\subsubsection{Eigenvalues at $k=0$ and correspondence with string
ground states}

The numerical results according for Fig.\ \ref{f:f5} imply a trend
that, towards the end of the tachyon condensation, there are four
degenerate zero modes under BC(1,2,4) (BC(2) contributes two). At
the end of the decay, a stable solution consisting of an
alternating series of D$p-1$ and anti-D$p-1$ branes form. It is
interesting to compare the four zero modes of the tachyon field at
$k=0$ to the bosonic massless states of the open string spectrum.

Without loss of generality, we assume that
the $x$ direction is compactified on a circle with radius
$R=1/\beta=\sqrt{2}$. The masses of open
strings stretched between the parallel branes are
\begin{equation}
   m^2=N+\left(\frac{\Delta{\theta} R}{2\pi}\right)^2-a,
\end{equation}
where $N$ is the level number,  $\triangle{\theta}$ is the angle
difference between any two branes and we recall that $\alpha'=1$.
The parameter $a$ is the normal ordering constant which is $0$ for
the Ramond (R) sector and $1/2$ for the Neveu-Schwarz (NS) sector.
Between the two branes located on a circle, there are six ways to
attach open strings: both ends on (i) D or (ii)
$\overline{\textrm{D}}$ branes; stretching from D to
$\overline{\textrm{D}}$ on both (iii) the right sides and (iv) the
left sides of the branes; stretching from $\overline{\textrm{D}}$
to D on both (v) the right sides and (vi) the left sides. The
bosonic massless states for the cases (i) and (ii) are the first
excited states in the NS sector, which contain the transverse
fluctuations of the brane. For (iii), (iv), (v) and (vi), the
ground states are massless when $\Delta{\theta}=\pi$ and
$R=\sqrt{2}$. These states become tachyonic if $R < \sqrt 2$.

Hence we find a total of 6 bosonic massless states. Two are concerned
with translations of the branes,   
two of them should be associated with brane-antibrane annihilation to D$p-3$
branes and the appearance of 
a new effective field theory \cite{Garousi:2004rd}, 
and the other two can be identified with the branes ``melting'' back 
into the unstable D$p$-brane.
It is interesting that the original tachyon field,
despite being singular, still contains sensible zero
modes.

\section{Motion in moduli space}
\label{sec:motion}

As we have seen, the condensation procession is essentially
described by any one of the following parameters: $T_0$,
$V_0=V(T_0)$ and $k=V_0/V_m$. In the following, we study the
time-dependent decay of the brane into a brane and an anti-brane
by studying the motion of the $k$ modulus based on the solution of
the static equation of motion.

We start from the $1+1$ dimensional effective action:
\begin{equation}
   S=-\int dtdx V(T) \sqrt{1-\dot{T}^{2}-T'^{2}}.
\end{equation}
In the moduli space, we adopt the approximation
$\dot{T}^{2}/(1+T'^{2})\ll 1$, i.e., the tachyon evolves very
slowly. Then an approximate action can be obtained, noting that
$T'^{2}=(V/V_0)^{2}-1$:
\begin{equation}
\label{e:Sapp}
   S\simeq -\int dtdx \frac{V^2}{V_0} \left[1-\frac{1}{2}\left(\frac{V_0}{V}\right)^2\dot{T}^{2}\right].
\end{equation}
When the second term is small enough, the Langrangian is exactly
the energy of the static case in Eq.\ (\ref{e:En}). We assume that
there is a slow-motion solution with the approximate form of Eq.\
(\ref{e:Tsol}), with $k=V_0/V_m$ time-dependent. Hence
\begin{equation}
   \dot{T}=-\frac{\tanh{(\beta T)}}{\beta k(1-k^2)}\dot{k}.
\end{equation}
Substituting into Eq.\ (\ref{e:Sapp}) and performing the
integration over $x$, over a complete period of the tachyon field,
we find
\begin{equation}
   S=\int dt L(k,\dot{k})= -\frac{2\pi V_m}{\beta}\int dt
\left[1-\frac{\dot{k}^2}{2\beta^2 k(1+k)(1-k^2)}\right].
\end{equation}
where $L(k,\dot{k})$ is the Langrangian with parameters of $k$ and
$\dot{k}$. The second term is a kinetic term, which can be
expressed in terms of a variable $X(t)$ defined through:
\begin{equation}
\frac{1}{2}\dot{X}^2=\frac{1}{2}g(k)\dot{k}^2,
\end{equation}
where, $g(k)={1}/{[\beta^2 k(1+k)(1-k^2)]}$. The solution of $X$
is trivial: $X=X_0+vt$, where $X_0$ and $v$ are constants.
Therefore from the relation between $k$ and $X$:
\begin{equation}
X=\int \sqrt{g(k)}dk=2\sqrt{\frac{\pi V_m}{\beta^3}}\tan^{-1}
\sqrt{\frac{2k}{1-k}},
\end{equation}
and we obtain the solution:
\begin{equation}
k(t)=\frac{4}{3+\cos{(c_1 t+c_0)}}-1,
\end{equation}
where $c_1=v\sqrt{\beta^3/\pi V_m}$ and $c_0$ are constants. We
discover that the D$p$ branes decay into D$p-1$ branes in a finite
time $\triangle t=\sqrt{V_m \pi^3/\beta^3}/v$. This result is
consistent with singularities appearing in finite time in
time-dependent tachyon simulations
\cite{Cline:2003vc,Barnaby:2004dz} and in boundary conformal field
theory calculations \cite{Sen:2002vv}. We note that the solution
is in fact periodic, and oscillates between the unstable D$p$
brane system and the daughter D$p-1$, $\overline{\textrm{D}}p-1$
brane system.

\section{Conclusion}
\label{sec:conclusions}

We have calculated numerically the low-lying modes of small
fluctuations in the tachyon field around the static solution
representing a periodic array of D$(p-1)$ and anti-D$(p-1)$ brane
pairs. The potential for the fluctuations can be expressed in
terms of Jacobi elliptic functions, and eigenvalue equations were
solved numerically.

We find the two expected zero modes, corresponding to changes in
the amplitude of the tachyon field between the kinks, and the
position of the kinks. In addition there is a negative eigenvalue
reflecting an attractive force between kinks and antikinks, which
causes them to move together and annihilate.

There are higher excited states with eigenvalues $\om^2
>\beta^2/4$. In the language of the effective field theory these
are unbound scattering states. In the limit that the tachyon field
between the D$(p-1)$-branes tends to the vacuum, the eigenvalues
of these states appear to tend to constant values (with $\om^2$
still greater than $\beta^2/2$), and the support of the
eigenfunctions shrinks with the width of the branes. Thus,
although propagating modes are present, they disappear from the
bulk between branes. This is consistent with the absence of
propagating modes in the case of homogeneous tachyon condensation.

We have also attempted to make the possible link of the degenerate
lowest tachyon fluctuation modes at the decay end $k=0$ to the
open string spectra of ground states in the daughter brane,
anti-brane system. With the daughter brane and anti-brane
compactified on a circle along the $x$ direction, we found that
the ground states of the open superstrings stretching between them
are indeed massless. But there are 6 such states, which are two more
than the zero modes we have obtained. The two extra zero massless 
states should correspond to standard brane-antibrane annihilation.

Finally, we note that there are approximate time-dependent
solutions where the parameter of the solution Eq.\ (\ref{e:Tsol}),
$V_0$ and $x_0$, become time-dependent. The time evolving
solutions show that there are singularities appearing in a finite
time, which is consistent with earlier work, but our solutions can
be continued through the singularity to produce a periodic
oscillation between a D$p$-brane and a D$p-1$-$\bar{\rm
D}p-1$-brane pair.

\acknowledgments We are very grateful to Koji Hashimoto for
correcting the discussion of the spectrum of the brane-antibrane
system. H. Li is supported by the Dorothy Hodgkin Postgraduate
Awards (DHPA).

\bibliographystyle{JHEP}

\bibliography{b}

\end{document}